\def\puncspace{\ifmmode\,\else{\ifcat.\C{\if.\C\else\if,\C\else\if?\C\else%
\if:\C\else\if;\C\else\if-\C\else\if)\C\else\if/\C\else\if]\C\else\if'\C%
\else\space\fi\fi\fi\fi\fi\fi\fi\fi\fi\fi}%
\else\if\empty\C\else\if\space\C\else\space\fi\fi\fi}\fi}
\def\SP{\let\\=\empty\futurelet\C\puncspace}

\def\h1{$h^{-1}$\SP}

\def\etal{{\it et al.\/}\ }

\def\void#1{{}}

\def\j{$J$}
\def\h{$H$}

\newcommand{\lsim}{\raisebox{-1.1mm}{$\stackrel{<}{\sim}$}}
\newcommand{\gsim}{\raisebox{-1.1mm}{$\stackrel{>}{\sim}$}}

\documentclass{aa}
\usepackage{graphicx}
\usepackage{times}
\usepackage{mathptm}
\fontfamily{ptmcm}\selectfont
\usepackage{graphicx}

\usepackage{psfig}

\include{psfig.sty}

\begin{document}
%\thesaurus{}

%\thesaurus{03(08.02.2, 08.04.1, 08.09.2, 11.13.1, 12.04.3)}

\title{ESO Imaging Survey}{\subtitle{Deep Public Survey: Infrared Data
for the Chandra Deep Field South\thanks{Based on observations
collected at the European Southern Observatory, La Silla, Chile within
program ESO 164.O-O561.}.}

\author{EIS-team}

%\void{
\author{B. Vandame\inst{1} 
 \and L.F. Olsen\inst{2}
 \and H.E. J{\o}rgensen\inst{2}	
 \and M.A.T. Groenewegen\inst{1}
 \and M. Schirmer\inst{1,3}
 \and S. Arnouts\inst{1}
 \and C. Benoist\inst{4,1}
 \and L. da Costa\inst{1}
 \and R. P. Mignani\inst{1}
 \and C. Rit\'{e}\inst{1,5}
 \and R. Slijkhuis\inst{1}
 \and E. Hatziminaoglou\inst{1}
 \and R. Hook\inst{1}
 \and R. Madejsky\inst{1}
 \and A. Wicenec\inst{1}
 }
%}
\institute{
European Southern Observatory, Karl-Schwarzschild-Str. 2, D--85748
Garching b. M\"unchen, Germany
\and Astronomical Observatory, Juliane Maries Vej 30, DK-2100 Copenhagen, 
Denmark 
\and Max-Planck Institut f\"ur Astrophysik, Karl-Schwarzschild-Str. 1, D-85748 Garching b.  M\"unchen, Germany
\and Observatoire de la C\^{o}te d'Azur, BP 229, 06304 NICE
cedex 4, France 
\and Observat\'orio Nacional, Rua Gen. Jos\'e Cristino 77, Rio de
Janerio, R.J., Brasil 
}

\offprints{bvandame@eso.org}

\date{Received ; accepted}

\abstract{This paper presents new $J$ and $K_s$ near-infrared data obtained from
observations of the Chandra Deep Field South (CDF-S) conducted at the
ESO 3.5m New Technology Telescope (NTT).  These data were taken as
part of the ongoing Deep Public Survey (DPS) being carried out by the
ESO Imaging Survey (EIS) program, extending the EIS-DEEP
survey. Combined these surveys now provide a contiguous coverage over
an area of 400~square~arcmin in the near-infrared, nearly matching
that covered by the deep X-ray observations of Chandra, four times the
area of the original EIS-DEEP survey.  The paper briefly describes the
observations and the new techniques being employed for pipeline
processing jittered infrared observations, which include unbiased
de-fringing and sky-background subtraction, pixel-based astrometry and
stacking and pixel registration based on a multi-resolution
decomposition of the images. The astrometric solution is based on a
pre-release of the GSC-II catalog and has an accuracy of
$\lsim0.15$~arcsec. The final images for 12 pointings presented here
reach median $5\sigma$ limiting magnitudes of $J_{AB}\sim23.4$ and
$K_{AB}\sim22.6$ as measured within an aperture 2$\times$FWHM. The
frame to frame variation of the photometric zero-point is estimated to
be $\lsim0.09$~mag.  The data are publicly available in the form of
fully calibrated $J$ and $K_s$ pixel maps and source lists extracted
for each pointing. These data can be requested through the URL
``http://www.eso.org/eis''.
\keywords{catalogs -- surveys-- stars: general - galaxies: general}
}

\maketitle

\section{Introduction}
\label{sec:intro}

\void{
Deep multi-wavelength imaging data of selected regions of the sky from
space and ground-based facilities combined with spectrographs on
large-aperture telescopes are}

Deep multi-wavelength observations of selected regions of the sky from
space and ground-based facilities, combined with spectrographic
observations from large-aperture telescopes, offer the most promising
means to probe the distant universe and study in a comprehensive way
the evolution of galaxies and large-scale structures over a broad
interval of look-back time. Preliminary steps towards this goal are
already underway with the completion of deep X-ray observations of the
Chandra Deep Field South (CDF-S; $\alpha = 03^h32^m28^s$ and $\delta =
-27\degr48\arcmin30\arcsec$; Giacconi \etal 2000). These observations
will eventually be complemented by observations with HST, XMM and
SIRTF as well as VLT.

Foreseeing the need for ground-based multi-color data the Working
Group for public surveys at ESO recommended the ESO Imaging Survey
(EIS, Renzini \& da Costa, 1997) project to undertake deep,
optical/infrared observations of the HDF-S and CDF-S regions. Original
observations of these fields were conducted in 1998 (da Costa \etal
1998; Rengelink \etal 1998, to be superseded by  da Costa \etal
2001; Benoist \etal 2001a) and fully calibrated images, source
catalogs and high-redshift galaxy candidates were immediately made
public prior to the Science Verification of the first unit of the
VLT. However, these observations covered only a small fraction of the
field of view of Chandra, requiring the original coverage of EIS to be
enlarged.  Full coverage of the CDF-S $\lsim350$~square~arcmin field
and flanking regions in optical and infrared is one of the primary
goals of the Deep Public Survey (DPS) now being conducted by EIS.  DPS
consists of two parts: first, a deep, optical multi-passband ($UBVRI$)
survey reaching limiting magnitudes $m_{AB}\sim26$~mag covering three
distinct strips (DEEP1, DEEP2 and DEEP3). These stripes consist of
four adjacent pointings (denoted by a-d in decreasing order of right
ascension) of the wide-field imager (WFI), mounted on the ESO/MPG 2.2m
telescope at La~Silla, yielding an area of one square degree each;
second, the contiguous coverage in the infrared passbands $JK_s$ of
two distinct regions (DEEP-2c and DEEP-3a) 450~square~arcmin each. The
infrared part of DPS is a joint effort between the EIS team and an
external group of scientist from ESO member states. Further details
about DPS can be found at the URL "http://www.eso.org/eis".  The
primary goal of the survey is to produce a data set from which
statistical samples of galaxies can be drawn to study the large scale
structures at high redshift. These data should also be valuable for
many other areas of investigation, in particular for
cross-identification with X-ray sources detected from the deep X-ray
exposure of Chandra.

The purpose of the present paper is to describe and present the data
from the new infrared observations of the CDF-S region.  The data
reported here complement those obtained by the original EIS-DEEP
survey (Rengelink
\etal 1998), which  are being re-analyzed and will be presented in a
separate paper (Benoist \etal 2001a).
%In order to overcome these problems a  new software
%package to reduce jittered infrared observations, briefly described in
%the present paper, was developed for the EIS pipeline. 
%To produce a
%homogeneous data set the data for EIS-DEEP are being re-analyzed and will be presented in a
%separate paper (Benoist \etal 2001a) which will supersede the original
%one (Rengelink \etal 1998).  
Section~\ref{sec:observations} reviews
the overall observing strategy and describes the
observations. Section~\ref{sec:reductions} discusses the techniques
employed in the data reduction and the photometric and astrometric
calibration of the images. Section~\ref{sec:catalogs} presents the
photometric parameters of the sources detected on each image
corresponding to different pointings and passband. Due to time
constraints other products such as image mosaics, catalogs extracted
from them, optical/infrared color catalogs, statistical samples and
lists containing other targets of potential interest will be presented
elsewhere. Even though the goal of the present paper is not to
interpret the data, Section~\ref{sec:results} presents the results of
comparisons between the present observations and those of other
authors. This is done for the sole purpose of assessing the quality of
the astrometry and photometry of the present data set. Finally, a
brief summary is presented in Section~\ref{sec:summary}.

\section{Observations}
\label{sec:observations}

\begin{figure*}
%\centerline{\hbox{\psfig{figure=deep2c_nosusi_BW.ps,angle=0,width=10cm,clip=}}}
\caption{Relative positions of the area
with $UBVRI$ coverage (outer square) and the infrared data from
EIS-DEEP (inner bright square) and the current survey. It can be seen
that the infrared data forms a contiguous mosaic covering a
significant fraction of the optical survey area. It also nearly
coincides with the area covered by the deep observations of Chandra.}
\label{fig:coverage} \end{figure*}

Infrared observations in the near-infrared passbands $J$ and $K_s$
were obtained using the SOFI camera (Moorwood, Cuby \& Lidman 1998)
mounted on the New Technology Telescope (NTT) at La~Silla. SOFI is
equipped with a Rockwell 1024$^2$ detector that, when used together
with its large field objective, provides images with a pixel scale of
0.29 arcsec, and a field of view of about $4.9\times 4.9$ square
arcmin.  The infrared pointings were chosen to produce a contiguous
coverage of a $4\times4$ mosaic when combined with those of
EIS-DEEP. They were also required to produce sufficient overlap
between adjacent pointings to enable the construction of an image
mosaic, avoiding severe decrease in the effective exposure time due to
the jitter pattern used and enabling a common photometric zero-point
to be determined. The inner $2\times2$ subset correspond to the
EIS-DEEP pointings reported by Benoist \etal (2001a). The remaining 12
pointings are those observed as part of the present survey which are
listed in Table~\ref{tab:pointings}. The table gives: in column (1)
the identification of each tile in the mosaic represented by the
indices $(i,j)$, where $i$ decreases with declination and $j$
decreases with right ascension; and in columns (2) and (3) the J2000
coordinates of the pointing. A schematic view of the coverage is
illustrated in Figure~\ref{fig:coverage} which shows the relative
positions of the region with $UBVRI$ coverage, the EIS-DEEP infrared
data and the newly accumulated data. In the convention described above
the pointings with (2,2) (2,3), (3,2) and (3,3), correspond to those
obtained by the EIS-DEEP program. The original goal of the
observations was to reach $5\sigma$ magnitudes of $J_{AB}$=23.5 and
$K_{AB}$=23.1 requiring a total integration time of 1800~sec and
7000~sec in $J$ and $K_s$, respectively, as estimated from the SOFI
exposure time calculator. Following a Working Group (WG)
recommendation, the total integration time of 8400~sec was later split
into 3600~sec in $J$ and 4800~sec in $K_s$ to reach $(J-K_s)\sim2$
(Vega system) at the limiting magnitude of the survey. The $J$
exposures were taken as a sequence of 60 one-minute exposures, while
the $K_s$ observations were split into two sequences of 40 one-minute
exposures.  However, during the reductions images with very high noise
level and/or unexplained gradients in the background were detected and
rejected from further processing. Therefore, in some cases the
effective exposure time may vary with respect to those originally
planned (see Section~\ref{sec:results}). In other cases, additional
exposures were obtained for a pointing. These additional exposures
typically originate from aborted exposure sequences, which were later
re-observed in their entirety.

\begin{table}
\caption{Central position of the 12~SOFI fields observed as part of the
DPS infrared survey of CDF-S (DEEP2c).}
\label{tab:pointings}
\center
\begin{tabular}{lcc}
\hline
\hline
$(i,j)$ & $\alpha$ (J2000) & $\delta$ (J2000)\\
\hline
(1,1) & $03^h32^m57^s$ & $-27\degr41\arcmin45\arcsec$\\
(1,2) & $03^h32^m37^s$ & $-27\degr41\arcmin45\arcsec$\\
(1,3) & $03^h32^m17^s$ & $-27\degr41\arcmin45\arcsec$\\
(1,4) & $03^h31^m57^s$ & $-27\degr41\arcmin45\arcsec$\\
(2,1) & $03^h32^m57^s$ & $-27\degr46\arcmin10\arcsec$\\
(2,4) & $03^h31^m57^s$ & $-27\degr46\arcmin10\arcsec$\\
(3,1) & $03^h32^m57^s$ & $-27\degr50\arcmin35\arcsec$\\
(3,4) & $03^h31^m57^s$ & $-27\degr50\arcmin35\arcsec$\\
(4,1) & $03^h32^m57^s$ & $-27\degr55\arcmin00\arcsec$\\
(4,2) & $03^h32^m37^s$ & $-27\degr55\arcmin00\arcsec$\\
(4,3) & $03^h32^m17^s$ & $-27\degr55\arcmin00\arcsec$\\
(4,4) & $03^h31^m57^s$ & $-27\degr55\arcmin00\arcsec$\\
\hline
\hline
\end{tabular}
\end{table}

A total of 16~nights were allocated to the infrared part of DPS split
into three runs: one in period~64 and two in period~66. The
observations was carried out for both the DEEP2c and the DEEP3a
regions, and a summary of the observations carried out in the DEEP2c
region is given in Table~\ref{tab:logs}, which lists: in column (1)
the date of the observations; in column (2) the pointing as defined
above; in column (3) the filters used; and in columns (4) and (5) the
range and average seeing during the night as measured by the seeing
monitor at La~Silla.  The observations were conducted under fairly
good weather and seeing ($\lsim1.1$~arcsec) conditions, nearly all in
photometric nights, except for occasional light cirrus. Unfortunately,
the last run was cancelled due to technical problems with the NTT. At
the time of writing the $J$ pointing (4,4) had been observed in
service mode and will be made available as soon as possible.

\begin{table*}
\caption{Log of the observations for the infrared data for DEEP2c.}
\label{tab:logs}
\center
\begin{tabular}{lcccc}
\hline\hline
Date & Pointing & Filter & Seeing range & Mean seeing\\ & (i,j) & &
    (arcsec) & (arcsec)\\
\hline
31-01-2000 & (1,3) & $K_s$ & $-$ & $-$\\
01-02-2000 & (1,3) & $K_s$ & $-$ & $-$\\
	   & (1,2) & $K_s$ & $0.66-1.07$ & $0.90$\\
03-02-2000 & (1,3) & $J$   & $0.79-1.35$ & $0.97$\\
	   & (1,2) & $K_s$ & $0.60-0.92$ & $0.76$\\ 
04-02-2000 & (1,2) & $J$   & $0.68-1.37$ & $0.91$\\
	   & (4,3) & $K_s$ & $0.69-1.17$ & $0.87$\\
05-02-2000 & (4,3) & $J$   & $0.44-0.83$ & $0.60$\\
06-02-2000 & (4,3) & $K_s$ & $0.43-0.63$ & $0.52$\\
19-11-2000 & (4,2) & $K_s$ & $0.70-1.00$ & $0.83$\\
	   & (4,2) & $J$   & $0.68-1.17$ & $0.85$\\
	   & (1,1) & $J$   & $0.70-2.26$ & $1.14$\\
	   & (1,1) & $K_s$ & $0.56-2.07$ & $0.95$\\
20-11-2000 & (1,1) & $K_s$ & $0.54-0.96$ & $0.75$\\
	   & (2,1) & $K_s$ & $0.54-1.01$ & $0.77$\\
	   & (2,1) & $J$   & $0.61-1.04$ & $0.79$\\
	   & (3,1) & $J$   & $0.63-1.17$ & $0.81$\\
	   & (3,1) & $K_s$ & $0.69-1.71$ & $1.10$\\
21-11-2000 & (4,1) & $K_s$ & $0.56-1.11$ & $0.82$\\
	   & (1,4) & $K_s$ & $0.64-1.33$ & $0.87$\\
	   & (4,1) & $J$   & $0.60-1.98$ & $1.16$\\
	   & (1,4) & $J$   & $0.54-1.58$ & $0.80$\\
	   & (2,4) & $K_s$ & $0.72-1.52$ & $1.01$\\
22-11-2000 & (2,4) & $K_s$ & $0.50-1.19$ & $0.79$\\
	   & (2,4) & $J$   & $0.60-2.40$ & $1.14$\\
	   & (3,4) & $K_s$ & $0.57-1.26$ & $1.00$\\
	   & (3,4) & $J$   & $0.62-1.49$ & $0.94$\\
	   & (4,4) & $K_s$ & $0.53-0.86$ & $0.68$\\
\hline\hline
\end{tabular}
\end{table*}

The infrared observations were jittered relative to the centers given
above in order to estimate the sky level from the data themselves. The
procedure consists of a series of short exposures with small position
offsets from the target position.  The jitter strategy has been
implemented as a standard observing template (AutoJitter) for the SOFI
instrument.  Each exposure consisted of the average of six ten-second
sub-exposures. Using this template, offsets are generated randomly
within a square box of a specified size, chosen originally to be
40~arcsec, approximately 10\% of the SOFI detector field of view. This
value was later increased to 70~arcsec for the November 2000 run to
improve the background estimate. These offsets are constrained so that
all distances between pointings, in a series of 15 consecutive
pointings, are larger than 9~arcsec.

\section{Data Reduction}
\label{sec:reductions}

\subsection {Image Processing}

The infrared images were reduced using a new routine designed for the
EIS pipeline for automated reduction of optical/infrared images
(Vandame \etal 2001).  The routine produces fully reduced images and
weight-maps carrying out bias subtraction, flatfielding, de-fringing
and subtraction of the background, first-order pixel-based image
stacking (allowing for translation, rotation and stretching of the
image) and astrometric calibration. De-fringing, sky-subtraction and
elimination of image artifacts are done adopting a two-pass procedure:
first, a preliminary image is obtained by stacking a set of
sky-subtracted frames, with the background estimated using a running
median; second, objects are identified in this stacked image and masks
are created; 
%NEW
\void{and enlarged by a user-specified dilation factor
(typically three)} 
finally, the de-fringing/sky-subtraction step is
repeated with the object masks mapped to the jittered position of each
individual image. Deviant values are sigma-clipped and masked pixels
are neglected. More details about the package will be presented when
it is made publicly available (Vandame \etal 2001).

The primary reason for developing an entire new code was to improve on
the photometric accuracy of faint objects which had been achieved in
the earlier processing of the EIS-DEEP data.
\void{objects, noticed after the release
of the EIS-DEEP infrared data.} Further analysis of these data carried
out by the EIS team and by other users (Cimatti 1999; see also the
eclipse home page\footnote{http://www.eso.org/projects/aot/eclipse/}
and Devillard 1999 for more information) revealed a trend to
underestimate the flux of objects, in some cases of $\sim$0.2~mag at
faint magnitudes. 
\void{This effect was largely independent of the input
parameters used for {\it
jitter}\footnote{http://www.eso.org/projects/dfs/papers/jitter99/},
developed primarily for on-line reductions, but sensitive to the
characteristics of the field being analyzed.  } 
Another important
advantage of the new software package is that it produces weight-maps
which are critical for the construction of image mosaics.

\begin{figure*}
\frame{\hbox{\psfig{figure=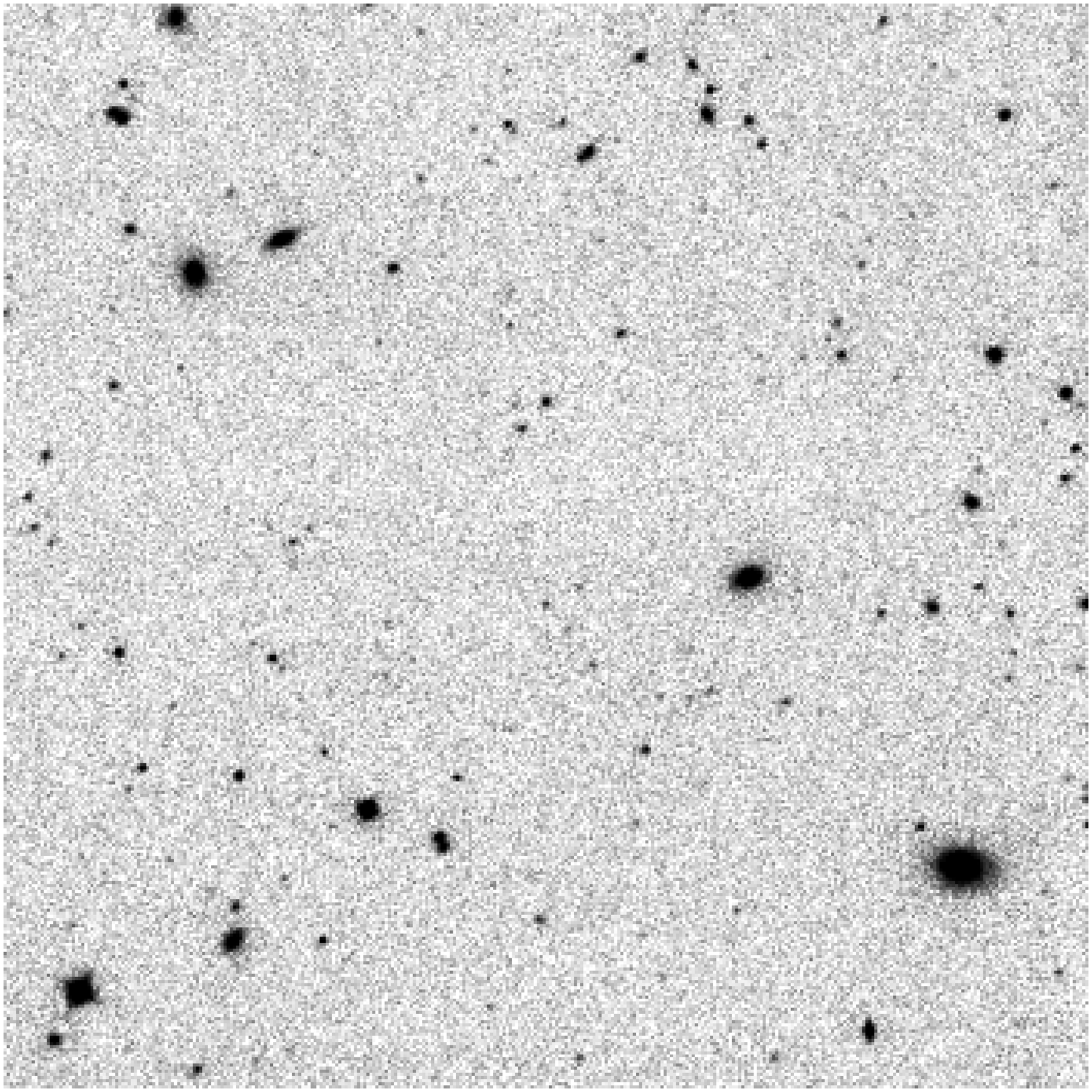,angle=0,width=\columnwidth,clip=}}}
\hspace{0.2cm}
\frame{\hbox{\psfig{figure=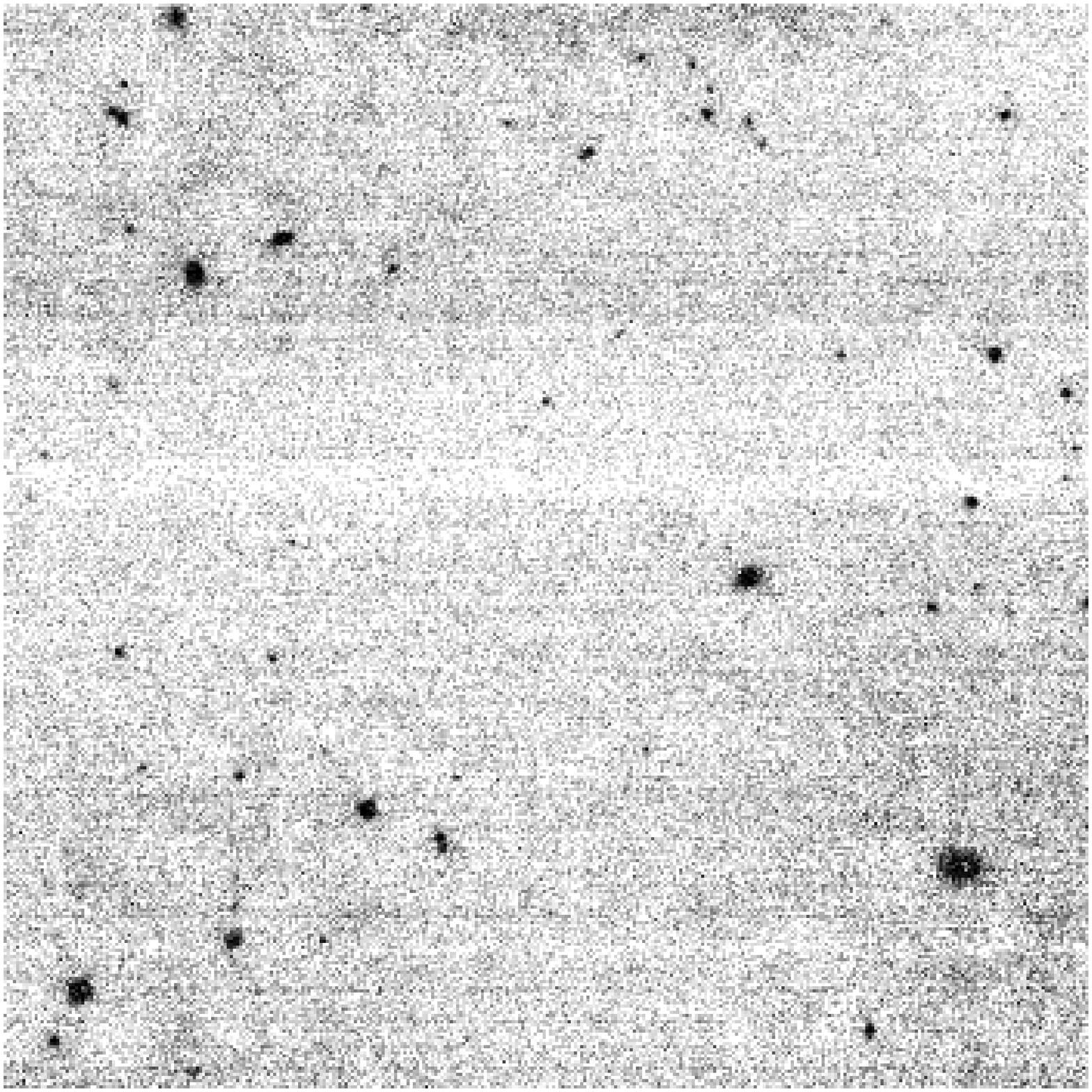,angle=0,width=\columnwidth,clip=}}}
\caption{Effect of the two-pass approach for the sky-estimation. The left
panel shows the final sky-subtracted image (note the inverted
fluxscale) after the second pass. The right panel shows the residuals
after subtracting the first pass image from the second pass one. The
large amounts of residual flux at the object positions indicate the
necessity for the second pass background estimation.}
\label{fig:irimage}
\end{figure*}

The effect of the two-step procedure is illustrated in
Figure~\ref{fig:irimage} which shows the sky-subtracted image
obtained after the two-pass sky-estimation (left panel) and the
residual after subtracting the first-pass image from the two-pass one
(right panel).  It can clearly be seen that a single pass leads to the
loss of a significant fraction of flux leading to biases in the
estimated magnitudes of up to $\lsim0.3$~mag.  This effect is caused
by the biased sky-estimation in the outer regions of the objects where
the faint light level contribution cannot be distinguished from the
sky-background in the individual images.  This effect is clearly
visible in the lower right corner where the background around the
bright galaxy is too high in the first pass estimate. Furthermore
regions with a high density of individual faint sources, none of which
can be detected prior to the final stacking, would contribute to the
background in the single pass approach.

\subsection {Astrometric Calibration}

The astrometric calibration of the infrared frames was performed using
as the reference catalog objects extracted from optical images
obtained over the same region (Arnouts \etal 2001) using the method
developed by Djamdji \etal (1993). The method is based on a
multi-resolution decomposition of images using wavelet transforms
(MVM) developed over the years by Bijaoui and collaborators and
extensively applied in remote sensing.  An implementation of these
algorithms to stack images, referred to as MVM-astrometry, has been
done for EIS.  This package has proven to be efficient and robust for
pipeline reductions. The estimated internal accuracy of this technique
is about half a pixel ($\sim$0.15~arcsec). The optical images were
calibrated relative to objects drawn from the pre-release version of
the {\em Guide Star Catalog-II} (GSC-II). The GSC-II (McLean \etal
 2001) is based on multi-passbands all-sky photographic plate
surveys including the Palomar Observatory Sky Survey (POSS-I-II), the
SERC and the ESO Red Survey. The astrometric calibration of GSC-II
has been obtained relative to Hipparcos (Perryman \etal 1997) and
Tycho catalogs (H{\o}g~\etal 1997) and the {\em ACT} (Urban, Corbin \&
Wycoff 1998).

This two-step approach to astrometrically calibrate the infrared
images has been taken for two main reasons.  First, by using the deep
astrometrically calibrated optical images as the reference the number
of objects typically available within a SOFI field ($\gsim 150$) is a
factor of four larger than the typical number of GSC-II reference
stars ($\sim40$). Therefore, this procedure provides a better
constraint for a second-order polynomial fit. Second, this approach
ensures a robust relative astrometry between the optical and infrared
images, overcoming the problems identified in the earlier release of
the CDF-S data.  As pointed out by Benoist \etal (2001b) small
relative shifts between the position of objects extracted from the
optical and infrared images significantly affected the color catalog
for that data set. The new procedure being adopted here should prevent
similar problems from occuring in the future.

\subsection{Photometric Calibration}

The photometric calibration for each night is based on observations of
standard stars taken from Persson \etal (1998). Typically two to seven
stars were observed over a range of airmasses. For all nights
independent photometric solutions were obtained with errors of
$\lsim~0.03$~mag in both passbands and negligible color
terms. Examining deviant measurements showed that some of the selected
standard stars (at least 3 cases) may be variable. A more careful
analysis of these cases will be presented elsewhere.

Using the individually calibrated images, the consistency between
their respective zero-points was examined considering the magnitude
differences of pairs of objects located in regions of overlap. An
average magnitude difference was then computed for each set of
overlapping frames yielding a mean value of $\sim~0.03$~mag and a
scatter of $\lsim0.09$~mag for both bands. A more detailed discussion
of this point will be presented in a forthcoming paper when the
complete data set has been analyzed and a final mosaic is created.

\subsection{Image Products}

The images being released are fully astrometrically (COE projection)
and photometrically calibrated (normalized to 1~sec exposure
time). The provided FITS files include both the pixel and weight maps
as FITS extensions.  The astrometric information is stored in the
world coordinate system (WCS) keywords in the FITS headers. Similarly
the photometric calibration is provided by the zero-point and its
error available in the header keywords ZP and ZP\_ERR. The zero-point
includes the normalized zero-point of the photometric solution and the
atmospheric extinction correction.  In the header one can also find a
product identification number (P\_ID}) which should always be used as
reference. The headers also provide information on the number of
stacked frames and the on-source total integration time.  Additionally
the seeing obtained by measuring the FWHM of bright stars in the final
stacked images is stored in the header keyword SEE\_IMA.

\begin{figure*}
%\frame{\hbox{\psfig{figure=k-img.eps,angle=0,width=\columnwidth,clip=}}}
\hspace{0.2cm}
%\frame{\hbox{\psfig{figure=k-weight.eps,angle=0,width=\columnwidth,clip=}}}
\caption{An example of the final coadded image (left panel) and the
corresponding weight map (right panel). The lines seen in the weight
map are caused by the astrometric remapping of the frames.}
\label{fig:imageK}
\end{figure*}

An example of a final image and corresponding weight map is shown in
figure~\ref{fig:imageK} for one pointing in $K_s$. Note that the size
of the image ($\sim5.8\times5.8$~square arcmin) is much larger than
that obtained using the earlier version of the EIS pipeline which
clipped the image to the best sampled region. Comparing the image and
the weight map one sees how the noise increases at the edges as the
effective exposure time decreases. However, proper use of the weight
maps facilitates and greatly increases the efficiency of building up
infrared mosaics.

It is worth mentioning that, a satellite track visible at the
north-eastern corner in one of the $J$-band images (pointing $(2,4)$)
led to the masking of a large region which caused the rejection of a
large fraction of the frame from the final stack. Proper use of the
Hough transform, as currently done for the optical images, will be
used to correct this effect in future data products. 
%(Benoist \etal, 2001b).

\begin{table*}
\caption{CDF-S \j-band source list (all magnitudes are given in the
$AB$ system).}
\centerline{\hbox{\psfig{figure=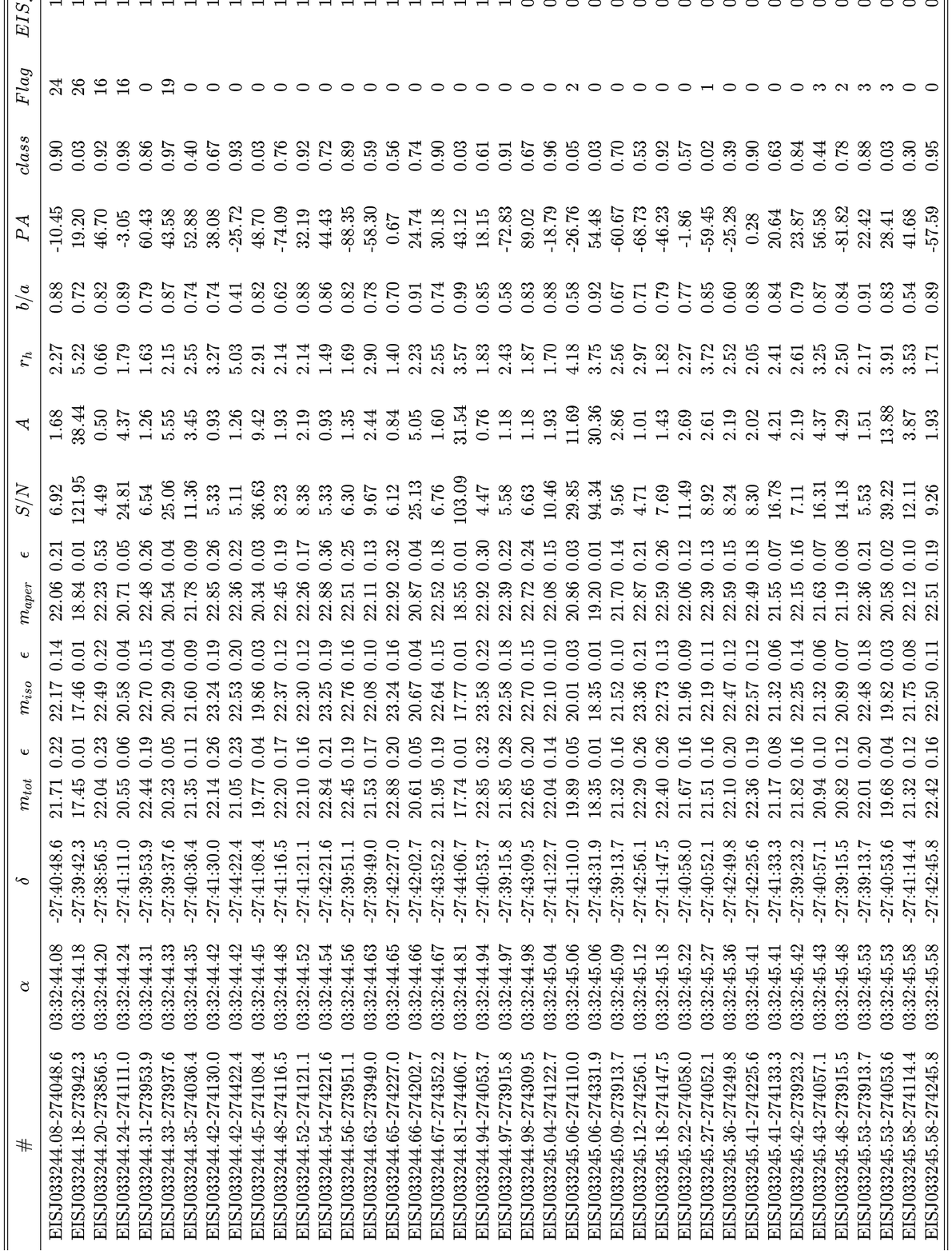,angle=0,width=\textwidth,clip=}}}
\label{fig:catalog}
\end{table*}

\section{Source Lists}
\label{sec:catalogs}

Source extraction was performed using the SExtractor software (Bertin
\& Arnouts 1996; ver. 2.2.1). Detection was carried out separately
using the co-added image of each passband and field.  The main
parameters in the detection are the smoothing kernel, taken to be a
Gaussian with a FWHM equal to 0.8 of that of the PSF measured on the
frame; the SExtractor detection threshold, taken to be 0.8; and the
minimum number of pixels above the detection threshold, taken to be 7
pixels for SOFI images. As an illustration, the tabulation of the
first 40 entries in one of the \j-band source catalogs is presented in
Table~\ref{fig:catalog}. All magnitudes are given in the $AB$ system,
using $J_{AB}= J + 0.90$ and $K_{AB}= K_s + 1.84$.  The table lists:

Column (1): the EIS identification name

Columns (2) and (3): right ascension and declination (J2000.0);

Columns (4)-(9): total, isophotal and aperture (3~arcsec diameter)
magnitudes and respective errors. The first two magnitudes correspond
to the {\tt mag\_auto} and {\tt mag\_iso} magnitudes measured by
SExtractor.  The magnitudes have been corrected for Galactic
extinction taken from Schlegel \etal (1998). The errors are those
estimated by SExtractor and include only the shot-noise of the
measured source and background counts. Only objects detected with
signal to noise $S/N\geq3$ (based on the isophotal magnitude errors)
are included.

Column (10): an estimate of the $S/N$ of the detection, by the inverse
of the errors estimated for the isophotal magnitude;

Columns (11): the isophotal area $A$ of the object in square arcsec;

Column (12): the half-light radius $r_h$  in arcsec;

Column (13) and (14): minor to major-axis  ratio and the position
angle;

Column (15): the stellarity index computed by SExtractor;

Column (16): SExtractor flags (see Bertin 1998)

Column (17): EIS flags; these flags are used to identify objects
in regions with very low weight or close to bright objects. Objects
detected in regions with weight $\geq40\%$ of the maximum weight
(proportional to the total integration time) and not affected by
bright stars have EIS flag=0. Objects with weight $<40\%$, located
at the edges of the frame, have EIS flag=1. Finally EIS flag=2
indicates that the object is located in a region masked out due to the
presence of a bright object.

Similar tables for each pointing and passband are available upon
request in ASCII format. Catalogs for the mosaic and optical/infrared
color catalogs will be distributed as soon as they become
available. Note that besides distributing catalogs in tabular form as
done here, catalogs in FITS formats will become part of the
distribution of final products. These catalogs will include
considerable more information such as the parameters defining the
trimmed and masked regions used to determine the EIS flags. The final
catalogs will be described by Arnouts \etal (2001).

\begin{table*}
\caption{Data summary. }
\label{tab:quality}
\center
\begin{tabular}{ccccccc}
\hline\hline
Pointing & filter & Eff.Exp.Time & Seeing & $N_{obj}$ & $5\sigma
 m_{lim}$ & $3\sigma m_{lim}$ \\ 
& & (sec) & (arsec) & & (mag) & (mag) \\
\hline
(1,1) & $J$   & 3600 & 0.99 & 662 & 23.39 & 23.94 \\ 
      & $K_s$ & 4800 & 0.84 & 530 & 22.43 & 22.98 \\ 
(1,2) & $J$   & 3600 & 0.77 & 736 & 23.48 & 24.03 \\ 
      & $K_s$ & 4800 & 0.76 & 795 & 22.54 & 23.09 \\ 
(1,3) & $J$   & 3600 & 0.84 & 549 & 23.23 & 23.78 \\ 
      & $K_s$ & 4680 & 0.69 & 722 & 22.82 & 23.37 \\ 
(1,4) & $J$   & 3600 & 0.77 & 871 & 23.59 & 24.14 \\ 
      & $K_s$ & 4500 & 0.77 & 663 & 22.64 & 23.19 \\ 
(2,1) & $J$   & 3600 & 0.83 & 699 & 23.47 & 24.02 \\ 
      & $K_s$ & 4740 & 0.68 & 654 & 22.60 & 23.15 \\ 
(2,4) & $J$   & 3600 & 0.82 & 534 & 23.02 & 23.57 \\ 
      & $K_s$ & 4800 & 0.81 & 566 & 22.43 & 22.98 \\ 
(3,1) & $J$   & 3600 & 0.69 & 896 & 23.85 & 24.40 \\ 
      & $K_s$ & 4800 & 0.84 & 442 & 22.40 & 22.95 \\ 
(3,4) & $J$   & 3600 & 0.87 & 697 & 23.47 & 24.02 \\ 
      & $K_s$ & 4800 & 0.77 & 602 & 22.62 & 23.17 \\ 
(4,1) & $J$   & 3600 & 1.03 & 627 & 23.14 & 23.69 \\ 
      & $K_s$ & 4800 & 0.68 & 602 & 22.63 & 23.18 \\ 
(4,2) & $J$   & 3720 & 0.86 & 588 & 23.38 & 23.93 \\ 
      & $K_s$ & 4920 & 0.85 & 415 & 22.37 & 22.92 \\ 
(4,3) & $J$   & 3600 & 0.66 & 810 & 23.42 & 23.97 \\ 
      & $K_s$ & 7200 & 0.77 & 633 & 23.02 & 23.57 \\ 
(4,4) & $K_s$ & 4800 & 0.67 & 710 & 22.71 & 23.26 \\ 
\hline\hline
\end{tabular}

\void{
\begin{tabular}{cccccccc}
\hline\hline
Tile  & filter & Seeing & $N_{obj}$ & $5\sigma m_{lim}$ & $N_{spur,5\sigma}$ & $3\sigma m_{lim}$ & $N_{spur,3\sigma}$\\
 & & (arsec) & & (mag) & \% & (mag) & \% \\
\hline
(1,1) & $J$   & 0.95 & 673 & 23.42 & & 23.97& \\ 
      & $K_s$ & 0.90 & & 22.28 & & 22.83& \\ 
(1,2) & $J$   & 0.75 & 752 & 23.50 & & 24.05& \\ 
      & $K_s$ & 0.72 & & 22.13 & & 22.68& \\ 
(1,3) & $J$   & 0.84 & 542 & 23.21 & & 23.76& \\ 
      & $K_s$ & 0.67 & & 22.48 & & 23.03& \\ 
(1,4) & $J$   & 0.78 & 852 & 23.57 & & 24.12& \\ 
      & $K_s$ & 0.75 & & 22.49 & & 23.04& \\ 
(2,1) & $J$   & 0.80 & 712 & 23.49 & & 24.04& \\ 
      & $K_s$ & 0.70 & & 22.21 & & 22.76& \\ 
(2,4) & $J$   & 0.85 & 512 & 22.96 & & 23.51& \\ 
      & $K_s$ & 0.80 & & 22.08 & & 22.63& \\ 
(3,1) & $J$   & 0.74 & 838 & 23.76 & & 24.31& \\ 
      & $K_s$ & 0.80 & & 22.16 & & 22.71& \\ 
(3,4) & $J$   & 0.90 & 679 & 23.42 & & 23.97& \\ 
      & $K_s$ & 0.90 & & 22.10 & & 22.65& \\ 
(4,1) & $J$   & 1.00 & 430 & 23.17 & & 23.72& \\ 
      & $K_s$ & 0.80 & & 22.07 & & 22.62& \\ 
(4,2) & $J$   & 0.90 & 572 & 23.32 & & 23.87& \\ 
      & $K_s$ & 0.85 & & 22.37 & & 22.92& \\ 
(4,3) & $J$   & 0.65 & 823 & 23.42 & & 23.97& \\ 
      & $K_s$ & 0.78 & & 22.42 & & 22.97& \\ 
(4,4) & $K_s$ & 0.72 & & 22.62 & & 23.17& \\ 
\hline\hline
\end{tabular}
}

\void{
\begin{tabular}{lccccccc}
\hline\hline
Pointing & Filter & Seeing & $N_{obj}$ &$5\sigma m_{lim}$ & $f$ &
$3\sigma m_{lim}$ & f \\
& & (arcsec) &  & (mag) &  &  &\\
\hline
(1,1) & $J$   & 0.95 & & 23.53\\
      & $K_s$ & \\
(1,2) & $J$   & 0.75 & & 23.61 \\
      & $K_s$ & \\
(1,3) & $J$   & 0.90 & & 23.25\\
      & $K_s$ & \\
(1,4) & $J$   & 0.75 & & 23.72\\
      & $K_s$ & \\
(2,1) & $J$   & 0.80 & & 23.61\\
      & $K_s$ & \\
(2,4) & $J$   & 0.85 & & 23.08\\
      & $K_s$ & \\
(3,1) & $J$   & 0.80 & & 23.79\\
      & $K_s$ & \\
(3,4) & $J$   & 0.90 & & 23.53\\
      & $K_s$ & \\
(4,1) & $J$   & 1.00 & & 23.28\\
      & $K_s$ & \\
(4,2) & $J$   & 0.90 & & 23.43\\
      & $K_s$ & \\
(4,3) & $J$   & 0.65 & & 23.53\\
      & $K_s$ & \\
(4,4) & $K_s$ & \\
\hline\hline
\end{tabular}
}
\end{table*}

\section{Discussion}
\label{sec:results}

The characteristics of the data obtained by the present infrared
survey are summarized in Table~\ref{tab:quality} which lists: in
column (1) the identification of the pointing; in column (2) the
filter; in column (3) the effective exposure time for each final
image; in column (4) the seeing of the final co-added image; in column
(5) the number of detected objects with $S/N\geq3$ as described above;
in columns (6) and (7) the $5\sigma$ and $3\sigma$ limiting $AB$
magnitude measured within an aperture $2\times$FWHM. At the $5\sigma$
limiting magnitudes the fraction of spurious objects is estimated to
be $\sim5\%$ and $\sim11\%$ in $J$ and $K$, respectively, rapidly
rising to over 50\% at the $3\sigma$ limit.  The fraction of spurious
objects was estimated by creating catalogs from the survey images
multiplied by $-1$. The mock images were then used as input to the
catalog production and from the comparison between the catalogs of the
mock and real objects the fraction of false-positives was estimated.

\begin{figure}
\centerline{\hbox{\psfig{figure=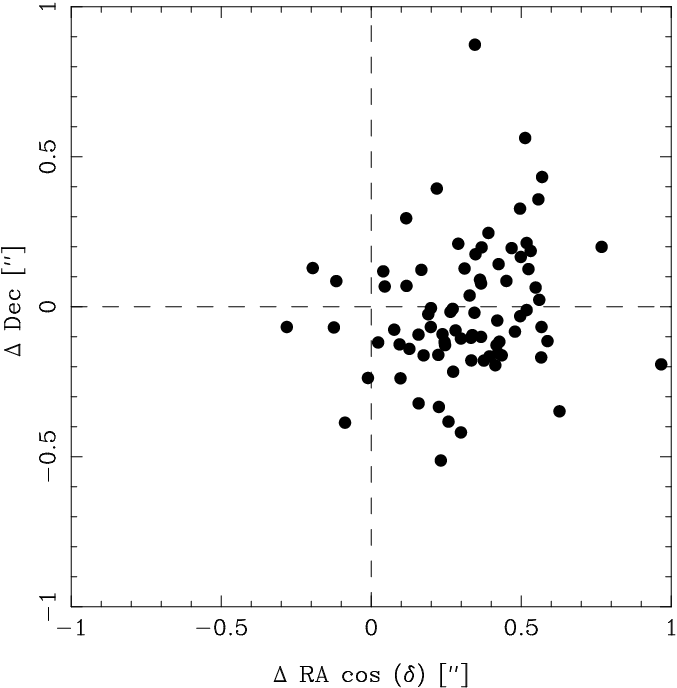,angle=0,width=\columnwidth,clip=}}}
\caption{Comparison between the location of objects in common with the
2MASS survey, offsets are computed as $EIS-2MASS$.}
\label{fig:astro_2mass} \end{figure}

\begin{figure}
\centerline{\hbox{\psfig{figure=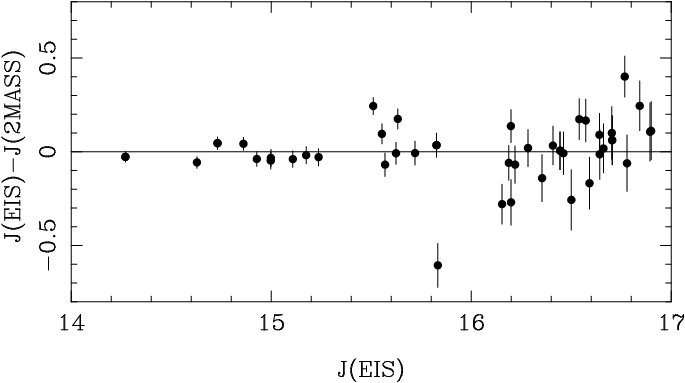,angle=0,width=\columnwidth,clip=}}}
\vspace{0.2cm}
\centerline{\hbox{\psfig{figure=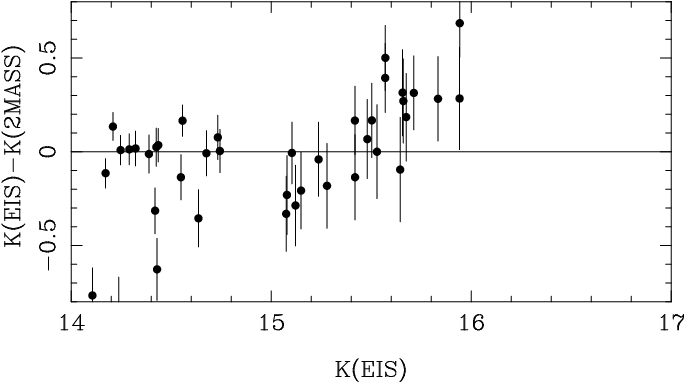,angle=0,width=\columnwidth,clip=}}}
\caption{Comparison between the measured $J$ (upper panel) and $K_s$
(lower panel) magnitudes of objects in common with the 2MASS survey.}
\label{fig:phot_2mass} \end{figure}

In order to externally verify the results obtained here the source
list presented above was cross-correlated with that available from the
2MASS survey (Cutri \etal 2000) using a search radius of 1~arcsec. A
total of 80 objects were found in common and their relative positions
and magnitudes (for objects considered ``good'' in both catalogs) were
computed and are shown in Figures~\ref{fig:astro_2mass} and
\ref{fig:phot_2mass}. Apart from an offset of about 0.33~arcsec in
right ascension one finds good agreement between the astrometric
calibrations yielding a scatter of $\lsim0.23$~arcsec. Assuming
similar errors for both data sets one finds an astrometric accuracy of
$\sim0.16$~arcsec consistent with the estimated internal accuracy of
the technique adopted in the present work. The asymmetry of the
distribution reflects a small systematic increase of the right
ascension residuals noted near the eastern edge of the region. This
effect is likely due to the large offset of the CDF-S position
relative to the center of the original plate used for GSC-II object
extraction. This problem is being addressed for the final release of
the catalog.  Considering objects brighter than $J=16.0$ and $K_s=15.0$
(Vega system), roughly corresponding to the 2MASS limiting magnitudes, the
measured magnitudes of the two data sets are in good agreement, except
for three outliers, showing a mean offset of 0.016~mag in $J$ and
$-0.031$~mag in $K_s$. The scatter is found to be 0.077~mag and
0.14~mag, respectively, consistent with the estimated uncertainty of
the overall zero-point (Section~\ref{sec:reductions}).

\begin{figure}
\centerline{\hbox{\psfig{figure=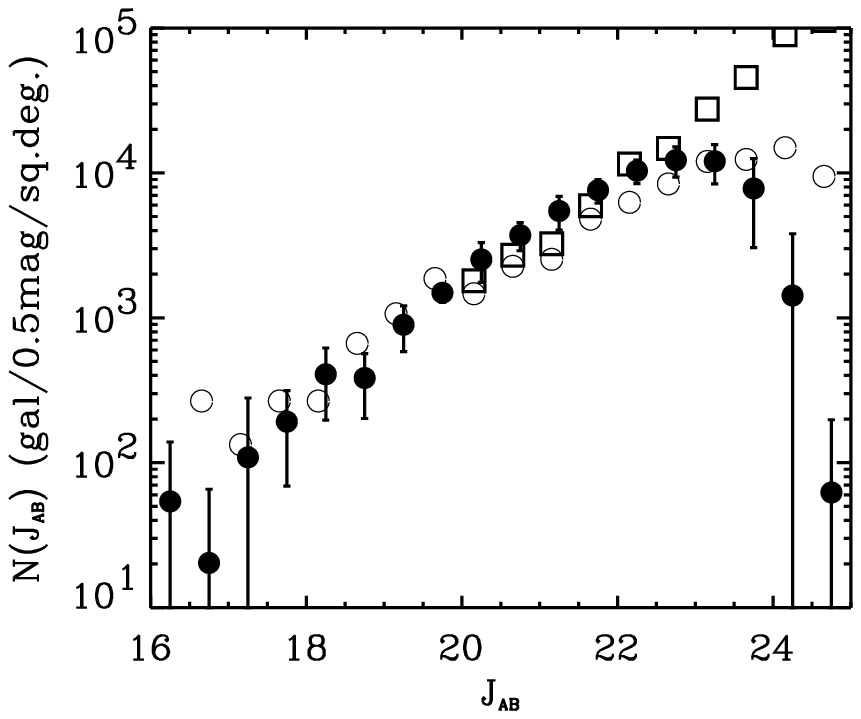,angle=0,width=\columnwidth,clip=}}}
\centerline{\hbox{\psfig{figure=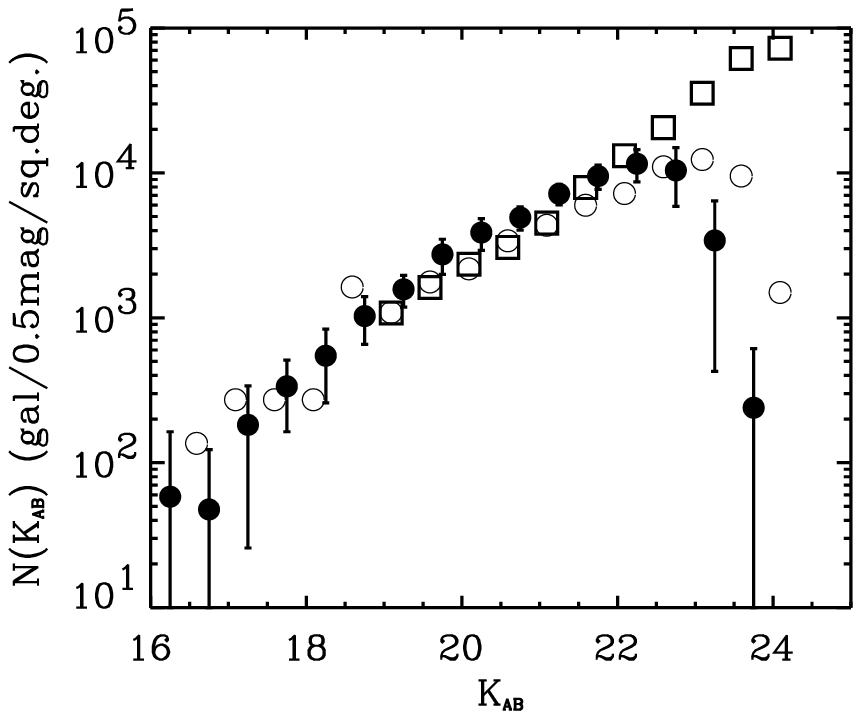,angle=0,width=\columnwidth,clip=}}}
\caption{Comparison of galaxy number counts between: the
present work (full circles); da Costa \etal (1998) obtained from
EIS-DEEP (open circles) and those of Saracco \etal (1999, squares).}
\label{fig:counts} \end{figure}

Finally, Figure~\ref{fig:counts} shows the comparison of galaxy counts
as a function of the total magnitude with those obtained for other
data sets.  The derived slopes of the counts are $0.33 \pm 0.02$ in
$J$ and $0.35 \pm 0.02$ in $K$, in good agreement with those obtained
by Saracco \etal (1999) as well as other studies. The galaxy counts
shown in the figure is the mean counts obtained averaging all the
individual catalogs and the error bar is the scatter. The individual
catalogs were obtained drawing objects with stellarity index less than
0.95 or fainter than 20.5~mag and only including objects with
$S/N>3$. Bad objects were discarded based on the SExtractor and EIS
flags.

\void{
\begin{figure}
\centerline{\hbox{\psfig{figure=spur_plot_J.ps,angle=0,width=\columnwidth,clip=}}}
\centerline{\hbox{\psfig{figure=spur_plot_K.ps,angle=0,width=\columnwidth,clip=}}}
\caption{Fraction of spurious objects}
\label{fig:spurious} \end{figure}
}

\section{Summary}
\label{sec:summary}

This paper presents the first results of the infrared data accumulated
by the ongoing Deep Public Survey being conducted by the EIS
program. The paper describes the observations and the first set of
products being publicly released including fully calibrated pixel maps
and source lists covering the CDF-S region. Given the current interest
in this area of the sky it was decided to release the data in the
simple format presented. Other derived products will be described in
forthcoming papers of this series. Hopefully, the source lists
presented here will be of immediate value for cross-identification
with X-ray sources.

\acknowledgements
The Guide Star Catalogue-II is produced by the Space Telescope Science
 Institute  in  collaboration   with the Osservatorio   Astronomico di
 Torino.    Space Telescope  Science   Institute  is  operated  by the
 Association    of Universities for  Research   in  Astronomy, for the
 National   Aeronautics    and  Space  Administration   under contract
 NAS5-26555.  Additional support  is  provided by the Association   of
 Universities  for Research   in  Astronomy, the  Italian Council  for
 Research in Astronomy, European Southern Observatory, Space telescope
 European Coordinating  Facility, the International GEMINI project and
 the European  Space Agency Astrophysics  Division.  The UK Schmidt
 Telescope  was operated by   the  Royal Observatory  Edimburgh,  with
 funding  from the UK Science  and Engineering Research Council (later
 the UK Particle Physics  and Astronomy Research Council), until  1988
 June, and thereafter by the Anglo-Australian Observatory \\
This publication makes use of data products from the Two Micron All
Sky Survey, which is a joint project of the University of
Massachusetts and the Infrared Processing and Analysis
Center/California Institute of Technology, funded by the National
Aeronautics and Space Administration and the National Science
Foundation.\\
We thank all of those directly or indirectly involved in the EIS
effort.  Our special thanks to M. Scodeggio for his continuing
assistance in a variety of issues, A. Bijaoui for allowing us to use
tools developed by him and collaborators over the years and past EIS
team members for building the foundations of this program.  We would
also like to thank A. Renzini and the members of the Working Group for
Public Surveys.


\begin{thebibliography}{}
\bibitem[]{arnouts:01} Arnouts, S. \etal, 2001, In preparation
\bibitem[]{benoist:01a} Benoist, C. \etal, 2001a, In preparation
\bibitem[]{benoist:01b} Benoist, C. \etal, 2001b, In preparation
\bibitem[]{berint:1996} Bertin, E. \& Arnouts, S.; 1996, A\&AS, 117, 393
\bibitem[]{bertin:97} Bertin, E., 1998, SExtractor, User's guide, v2.0
\bibitem[]{cimatti:99} Cimatti, A, 1999, Private communication
\bibitem[]{cutri:00} Cutri, R.M.; Skrutskie, M.F.; Van Dyk, S.;
% Chester T., Evans T., 
  	\etal, 2000, Explanatory Supplement to the 2MASS Second
Incremental Data Release
\bibitem[]{dacosta:01} da Costa, L. \etal, 2001, In preparation
\bibitem[]{dacosta:98} da Costa, L. \etal, 1998, astro-ph/9812105, submitted to A\&A
\bibitem[]{devillard:97} Devillard, N., 1997, The Messenger, 87, 19
\bibitem[]{devillard:99} Devillard, N, 1999, ADASS, 98, 333
\bibitem[]{djamdji:93} Djamdji, J.P.; Bijaoui, A. \& Mani\`{e}re, R.; 1993, 
        Photogrammetric Engineering and Remote Sensing, 59, 645
\bibitem[]{giaconni:00} Giaconni, R; Rosati, P.; Tozzi, P.; \etal, 2000,
astro-ph/0007240, submitted to ApJ
\bibitem[]{hoeg:97} H{\o}g, E.; B\"{a}ssgen, G.; Bastian, U.; \etal, 1997,
A\&A, 323, L57
\bibitem[]{mclean:00} McLean, B.;  Greene, G.; Lattanzi, M.; \etal,
2001, in Mining the Sky, Proc. of ESO Workshop, Springer-Verlag, In
press 
\bibitem[]{moorwood:98} Moorwood, A.;  Cuby, J.G.; \&  Lidman, C.;  1998, The Messenger, 91,9
\bibitem[]{perryman:97} Perryman, M. A. C.; Lindegren, L.; Kovalevsky, J.; \etal,
1997, A\&A, 323, L49
\bibitem[]{rengelink:98} Rengelink, R. \etal, 1998, astro-ph/9812190, submitted to A\&A
\bibitem[]{renzini:97} Renzini, A. \& da Costa, A., 1997, The
Messenger, 87, 23 
\bibitem[]{saracco:99} Saracco,P.; D'Odorico, S.; Moorwood, A.; \etal, 1999, A\&A,
349, 751
\bibitem[]{schlegel:98} Schlegel, D.; Finkbeiner, D. \& Davis,M.;
1998, ApJ, 500,525
\bibitem[]{urban:98} Urban, S. E.; Corbin, T. E. \& Wycoff, G. L.,  1998 AJ 115, 2161
\bibitem[]{vandame:01} Vandame, B. \etal, 2001b, In preparation


%GSC-II
%McLean, B.,  Greene, G., Lattanzi, M., Spagna, A., Carollo, D., Smart,
%R., Mignani. R.P, Bucciarelli, B. 2001, 
%
%Proc. of ESO Workshop on Mining the Sky, Springer-Verlag - in press 
%
%
%TYCHO:
%
%Hoeg, E.; Bässgen, G.; Bastian, U.; Egret, D.; Fabricius, C.;
% Großmann, V.; Halbwachs, J. L.; Makarov, V. V.;
% Perryman, M. A. C.; Schwekendiek, P.; Wagner, K.;
% Wicenec, A., 1997, A\&A 323, L57
%
%HIPPARCOS:
%
%Perryman, M. A. C.; Lindegren, L.; Kovalevsky, J.;
% Hoeg, E.; Bastian, U.; Bernacca, P. L.; Crézé, M.;
% Donati, F.; Grenon, M.; van Leeuwen, F.;
% van der Marel, H.; Mignard, F.; Murray, C. A.;
% Le Poole, R. S.; Schrijver, H.; Turon, C.; Arenou, F.;
% Froeschlé, M.; Petersen, C. S. 1997 A\&A 323, L49
%
%ACT= Astrographtyc Catalogue +Tycho
%
%Urban, S. E., Corbin, T. E., Wycoff, G. L.,  1998 AJ 115, 2161
%
%main article):
%  1.Geometrical Registration of images. The multiresolution approach.
%        Djamdji J.P., Bijaoui A., Manière R.:
%        Photogrammetric Engineering and Remote sensing 59 pp.645-653 1993.
%
%Cutri et al. 2000
%Cutri R.M., Skrutskie M.F., Van Dyk S., Chester T., Evans T., 
%et al., 2000, Explanatory Supplement to the 2MASS Second
%Incremental Data Release



\end{thebibliography}
\end{document}